# DETERMINATION OF THE COUPLING COEFFICIENT C(ν) FROM THE CALCULATIONS OF THE LOW-FREQUENCY RAMAN SPECTRUM IN PMMA

## M.A. Korshunov

*L.V. Kirensky Institute of Physics Siberian Branch of the RAS, 660036 Krasnoyarsk, Russia*

*E-mail: kors@iph.krasn.ru*

Using the atom-atom potentials method, we have calculated the low-frequency Raman spectra of molecular crystal model with elements of disorder (Poly-Methyl-Methacrylate) PMMA that has a flexible molecule. Based on the results the coupling factor C (ν) has been calculated. Anharmonicity of vibrations, electro-optical anharmonicity, and disorder were taken into account.

In the low-frequency Raman spectra and infrared spectra in studies of disordered materials (amorphous, glassy polymers) appears boson peak. Boson peak reflects the vibrational properties of the medium and is associated with excess vibrational density of states, which arises due to lack of long-range order and all phonon states may participate in the scattering of light. Such features range may occur due to the substantial anharmonicity of the vibrations. Carried out by the method of Dean [1] model calculations PMMA molecular structures that take into account the effect of disorder and anharmonicity.

For disordered systems, the intensity of the spectral lines of Raman scattering [2] is related to the density of states:

$$I(\nu) = g(\nu)C(\nu)[n(\nu)+1]/\nu,$$

Here [n(ν) +1] is the Bose-factor, g(ν) is the density of vibrational states, and C(ν) is the matrix element of the vibrational modes. In [3] found that the C (ν) is proportional to the change in the polarizability fluctuations. Calculating the g(ν) by the method of Dean and Raman spectrum, taking into account the intensity of the spectral lines obtained through the polarizability tensor can be found C (ν). When calculating the intensities taken into account anharmonicity of the vibrations and electro-optical anharmonicity. These results are consistent with the data of Refs. [4,5].

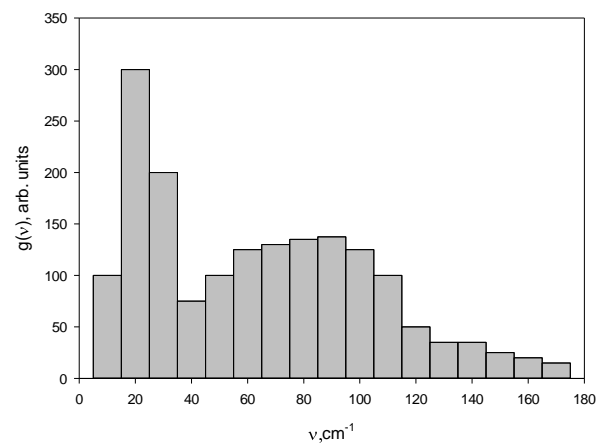

Figure 1. Histogram of the unordered PMMA spectrum at 300K with the disorder of the flexible molecules.

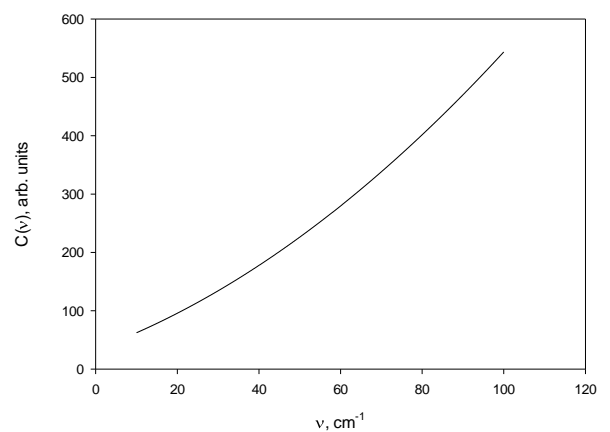

Figure 2. Calculated dependence of C (ν) on the frequency for PMMA at 300K.

Thus, calculations allows to immediately obtain a distribution of the density of vibrational states. As well, as the intensity of the lines, taking into account the anharmonicity of vibrations and electro-optical anharmonicity. This makes it possible to find the coupling coefficient $C(v)$.